\documentclass{aastex61}

\received{2017-10-15}
\revised{2017-10-29}
\accepted{2017-11-01}
\submitjournal{The Astrophysical Journal Letters}

\shorttitle{Dust Belts in Proxima Centauri}
\shortauthors{Anglada et al.}

\begin{document}

\title{ALMA Discovery of Dust Belts Around Proxima Centauri}

\correspondingauthor{Guillem Anglada}
\email{guillem@iaa.es}

\author{Guillem Anglada}
\affiliation{Instituto de Astrof\'{\i}sica de Andaluc\'{\i}a, CSIC, 
Glorieta de la Astronom\'{\i}a s/n, 18008 Granada, Spain}

\author{Pedro J. Amado}
\affiliation{Instituto de Astrof\'{\i}sica de Andaluc\'{\i}a, CSIC, 
Glorieta de la Astronom\'{\i}a s/n, 18008 Granada, Spain}

\author{Jose L. Ortiz}
\affiliation{Instituto de Astrof\'{\i}sica de Andaluc\'{\i}a, CSIC, 
Glorieta de la Astronom\'{\i}a s/n, 18008 Granada, Spain}

\author{Jos\'e F. G\'omez}
\affiliation{Instituto de Astrof\'{\i}sica de Andaluc\'{\i}a, CSIC, 
Glorieta de la Astronom\'{\i}a s/n, 18008 Granada, Spain}

\author{Enrique Mac\'{\i}as}
\affiliation{Department of Astronomy, Boston University, 725 
Commonwealth Avenue, Boston, MA 02215, USA}

\author{Antxon Alberdi}
\affiliation{Instituto de Astrof\'{\i}sica de Andaluc\'{\i}a, CSIC, 
Glorieta de la Astronom\'{\i}a s/n, 18008 Granada, Spain}

\author{Mayra Osorio}
\affiliation{Instituto de Astrof\'{\i}sica de Andaluc\'{\i}a, CSIC, 
Glorieta de la Astronom\'{\i}a s/n, 18008 Granada, Spain}

\author{Jos\'e L. G\'omez}
\affiliation{Instituto de Astrof\'{\i}sica de Andaluc\'{\i}a, CSIC, 
Glorieta de la Astronom\'{\i}a s/n, 18008 Granada, Spain}

\author{Itziar de Gregorio-Monsalvo}
\affiliation{European Southern Observatory, 3107 Alonso de C\'ordova, 
Vitacura, Santiago, Chile}
\affiliation{Joint ALMA Observatory, Alonso de C\'ordova 3107, Vitacura, 
Casilla 19001, Santiago 19, Chile}

\author{Miguel A. P\'erez-Torres}
\affiliation{Instituto de Astrof\'{\i}sica de Andaluc\'{\i}a, CSIC, 
Glorieta de la Astronom\'{\i}a s/n, 18008 Granada, Spain}
\affiliation{Visiting Scientist: 
Departamento de Física Te\'orica, Facultad de Ciencias, Universidad de 
Zaragoza, E-50009, Zaragoza, Spain}

\author{Guillem Anglada-Escud\'e}
\affiliation{School of Physics and Astronomy, Queen Mary University of 
London, 327 Mile End Rd., London E1 4NS, UK}

\author{Zaira M. Berdi\~nas}
\affiliation{Departamento de Astronom\'{\i}a, Universidad de Chile, Camino 
del Observatorio, 1515 Las Condes, Santiago, Chile}
\affiliation{Instituto de Astrof\'{\i}sica de Andaluc\'{\i}a, CSIC, 
Glorieta de la Astronom\'{\i}a s/n, 18008 Granada, Spain}

\author{James S. Jenkins}
\affiliation{Departamento de Astronom\'{\i}a, Universidad de Chile, Camino 
del Observatorio, 1515 Las Condes, Santiago, Chile}

\author{Izaskun Jimenez-Serra}
\affiliation{School of Physics and Astronomy, Queen Mary University of 
London, 327 Mile End Rd., London E1 4NS, UK}

\author{Luisa M. Lara}
\affiliation{Instituto de Astrof\'{\i}sica de Andaluc\'{\i}a, CSIC, 
Glorieta de la Astronom\'{\i}a s/n, 18008 Granada, Spain}

\author{Maria J. L\'opez-Gonz\'alez}
\affiliation{Instituto de Astrof\'{\i}sica de Andaluc\'{\i}a, CSIC, 
Glorieta de la Astronom\'{\i}a s/n, 18008 Granada, Spain}

\author{Manuel L\'opez-Puertas}
\affiliation{Instituto de Astrof\'{\i}sica de Andaluc\'{\i}a, CSIC, 
Glorieta de la Astronom\'{\i}a s/n, 18008 Granada, Spain}

\author{Nicolas Morales}
\affiliation{Instituto de Astrof\'{\i}sica de Andaluc\'{\i}a, CSIC, 
Glorieta de la Astronom\'{\i}a s/n, 18008 Granada, Spain}

\author{Ignasi Ribas}
\affiliation{Institut de Ci\`encies de l'Espai (IEEC-CSIC), C/ Can Magrans s/n, 
Campus UAB, 08193 Bellaterra, Spain}

\author{Anita M. S. Richards}
\affiliation{JBCA, School of Physics and Astronomy, University of 
Manchester, Manchester, M13 9PL, UK}

\author{Cristina Rodr\'{\i}guez-L\'opez}
\affiliation{Instituto de Astrof\'{\i}sica de Andaluc\'{\i}a, CSIC, 
Glorieta de la Astronom\'{\i}a s/n, 18008 Granada, Spain}

\author{Eloy Rodriguez}
\affiliation{Instituto de Astrof\'{\i}sica de Andaluc\'{\i}a, CSIC, 
Glorieta de la Astronom\'{\i}a s/n, 18008 Granada, Spain}

%
%
%
%



\begin{abstract}

Proxima Centauri, the star closest to our Sun, is known to host at least 
one terrestrial planet candidate in a temperate orbit. Here we report 
the ALMA detection of the star at 1.3 mm wavelength and the discovery of 
a belt of dust orbiting around it at distances ranging between 1 and 4 
au, approximately. Given the low luminosity of the Proxima Centauri 
star, we estimate a characteristic temperature of about 40 K for this 
dust, which might constitute the dust component of a small-scale analog 
to our solar system Kuiper belt. The estimated total mass, including 
dust and bodies up to 50 km in size, is of the order of 0.01 Earth 
masses, which is similar to that of the solar Kuiper belt. Our data also 
show a hint of warmer dust closer to the star. We also find signs of two 
additional features that might be associated with the Proxima Centauri 
system, which, however, still require further observations to be 
confirmed: an outer extremely cold (about 10 K) belt around the star at 
about 30 au, whose orbital plane is tilted about 45 degrees with respect 
to the plane of the sky; and additionally, we marginally detect a 
compact 1.3 mm emission source at a projected distance of about 1.2 
arcsec from the star, whose nature is still unknown.

\end{abstract}





\section{Introduction} \label{sec:intro}

Cold debris disks around main sequence stars (e.g., Greaves et al. 2004; 
Di Folco et al. 2007; Lestrade et al. 2012; Chavez-Dagostino et al. 
2016; MacGregor et al. 2016) are left-over planetesimals that could not 
agglomerate into larger bodies during the process of planet formation. 
In these disks, dust grains are continuously replenished by dust 
particles resulting from grinding-down of larger planetesimals in 
destructive collisions (the so-called collisional cascade; Wyatt et al. 
2007a). These processes produce a second-generation of dust grains with 
a wide size distribution, whose thermal emission is observable from 
far-IR to mm wavelengths (Wyatt 2008; Matthews et al. 2014, and 
references therein). Dust is usually distributed as a belt within the 
periphery of the system in a way analogous to the Kuiper belt in our 
Solar System. While most of the known debris disks present cold dust in 
narrow belts at tens of au, a small fraction host a hot dust component 
within a few au, analogous to the Asteroid belt or Zodiacal dust (e.g., 
Absil et al. 2013; Marino et al. 2017).

The study of the present-day structure and dynamics of these dust belts 
can provide important information about the formation and evolution of 
exoplanetary systems, in a similar way as it has been done in our solar 
system. For example, it is thought that the present-day solar Kuiper 
belt is more extended and 100 times less massive than it was initially. 
This depletion in mass is explained in terms of dynamical instabilities  
due to early Jupiter-Saturn interactions (the Nice model; Morbidelli et 
al. 2005). 

Proxima Centauri, at a distance of 1.3 pc (van Leeuwen 2007), is the 
closest star to the Sun. It is an M5.5V star with $T_{\rm eff}=3000$~K, 
$M_*=0.12~M_\sun$, $R_*=0.15~R_\sun$, and $L_*=0.0015~L_\sun$ (Ribas et 
al. 2017); it belongs to a triple system (Kervella et al. 2017) and its 
age is estimated to be $\sim$5 Gyr (Bazot et al. 2016), if coeval 
formation is assumed. The discovery, using Doppler data, of a 
terrestrial planet candidate ($m_p \sin i=1.3~M_\earth$) orbiting the 
star at 0.05 au (Anglada-Escud\'e et al. 2016) has triggered the study 
of the main features of this stellar-planetary system.
 Raymond et al. (2011) show that debris disks are signposts of 
terrestrial planet formation. Thus, one might expect that Proxima 
Centauri is associated with a Kuiper belt analog that would allow us to 
learn about its planetary system history and architecture.
 Besides the intrinsic interest of studying a Kuiper belt analog in 
Proxima Centauri, imaging this belt would allow us to constrain the 
inclination angle of the orbital plane of the planet Proxima b and 
therefore to determine its true mass. 

In this Letter we report the first results of ALMA band 6 observations 
towards the star closest to our Sun, aiming at characterizing the 
architecture of its planetary system through the thermal emission of the 
surrounding dust.

\section{Observations} \label{sec:obs}

We observed Proxima Centauri with the Atacama Large 
Millimeter/Submillimeter Array (ALMA) at 1.3 mm, using both the Atacama 
Compact Array (ACA) of 7-m antennas and the main ALMA array of 12-m 
antennas.
 In all cases, we observed in dual polarization, with four spectral 
windows centered at 225, 227, 239 and 241 GHz, each with a 1.875 GHz 
effective bandwidth split into 120 channels. ACA observations were made 
between 2017 January 21 and 2017 March 24, with 8-11 antennas available, 
in 13 separate sessions of $\sim1.6$ hours each, including overheads. 
Ganymede, Callisto, Titan, J1427$-$4206, and J1517$-$2422 were used for 
absolute flux calibration, J1427$-$4206 and J1924$-$2914 for bandpass 
calibration, and J1424$-$6807 and J1329$-$5608 (within $5.5\degr$ and 
$10.0\degr$ of the target, respectively) for phase calibration. 
Observations were sensitive to angular scales $\la29''$. At this band, 
the primary beam FWHM is $\sim39''$ and its first null is at a radius of 
$\sim46''$. The 12-m array, with 41 antennas available, was used in a 
single session of 2.6 h on 2017 April 25. J1617$-$5848 and Titan were 
used for absolute flux calibration, J1427$-$4206 for bandpass 
calibration, and J1424$-$6807 (within $5.5\degr$ of the target) for 
phase calibration. Observations were sensitive to angular scales 
$\la6''$. The primary beam FWHM is $\sim23''$ and its first null is at a 
radius of $\sim27''$.

Considering the large proper motions and parallax of Proxima Centauri, 
the phase center of the observations was updated from scan to scan to 
track the source position with sub-milliarcsec accuracy. This was done 
by assuming source coordinates at epoch and equinox J2000.0 of 
R.A.=$14^h29^m42.9485^s$, Dec=$-62^\circ40'46.163''$, proper motions of 
$-$3775.75 and 765.54 mas yr$^{-1}$ (van Leeuwen et al. 2007) in R.A. 
and Dec, respectively, and a parallax of 768.13 mas (Lurie et al. 2014).

Data were processed using the Common Astronomy Software Applications 
(CASA) package. Calibration was performed with the ALMA pipeline, using 
CASA versions 4.7.0 and 4.7.2 for ACA and 12-m array data, respectively. 
Images were obtained by combining all spectral windows in multifrequency 
synthesis mode, applying a robust parameter of 0.5, and deconvolving 
with the CLEAN algorithm.

Self-calibration was attempted, but was unsuccessful due to insufficient 
S/N. Flux densities and intensities were measured with task IMSTAT. 
Source positions and sizes were determined with task IMFIT. Positional 
and flux uncertainties include absolute errors (5\% of the resolution in 
astrometry and 7\% in flux calibration) and relative errors due to noise 
in the images. All errors quoted in this Letter are at a $1\sigma$ 
level.

\begin{figure}[bt]
\begin{center}
\epsscale{0.8} 
\plotone{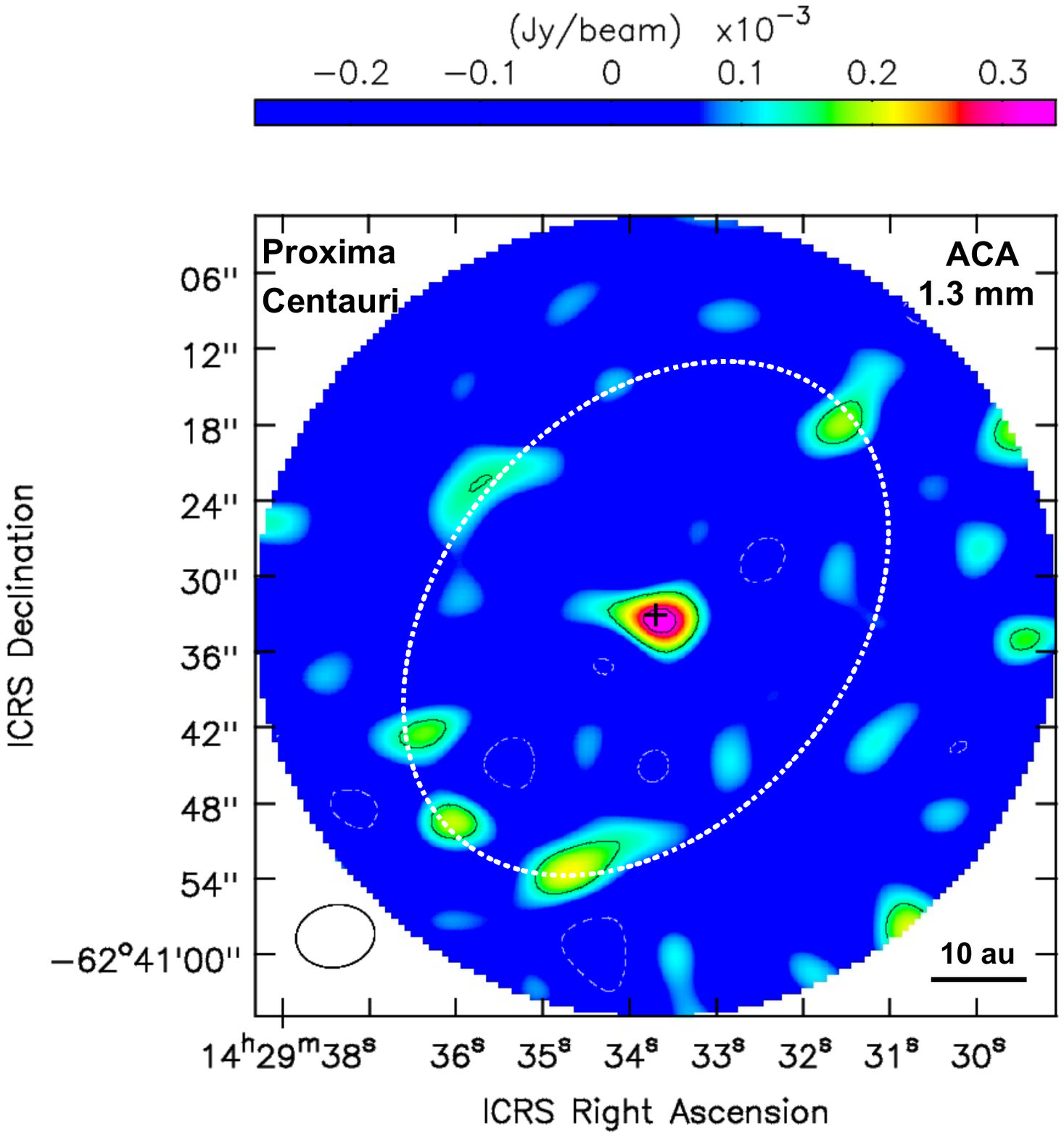}
 \caption{ACA 7-m array image of the 1.3 mm emission  
towards Proxima Centauri. Contours are $-$3, 3, and 6 times 50 $\mu$Jy 
beam$^{-1}$, the rms of the image.
 The position of the star on 2017 Jan 21, measured from an optical image 
(error $\simeq$ $0.1''$), is marked with a + sign. The ellipse shows the 
belt of radius 30 au suggested by the azimuthally averaged intensity 
profile (see text).
 The primary beam response correction has not been applied to this image 
to better represent the noise distribution. The figure shows the region 
where the primary beam response is $>20\%$ of that at the field center.
 The synthesized beam ($6.28''\times4.96''$, PA=$-80.4\degr$) is shown 
in the bottom-left corner.
 \label{fig:aca}}
\end{center}
 \end{figure}

Figure~\ref{fig:aca} shows the 1.3 mm ACA image (synthesized beam FWHM 
$\simeq$ $6''$). Since the phase center was changed in each session to 
track the source position, all data were assigned the formal position of 
the first epoch of observation, prior to combination of all epochs. Our 
image shows an unresolved source coinciding within the uncertainty 
($0.4''$) with the optical position of Proxima Centauri. The source has 
a flux density of $340\pm60$ $\mu$Jy.

\begin{figure}[tb]
\begin{center}
\epsscale{0.8} 
\plotone{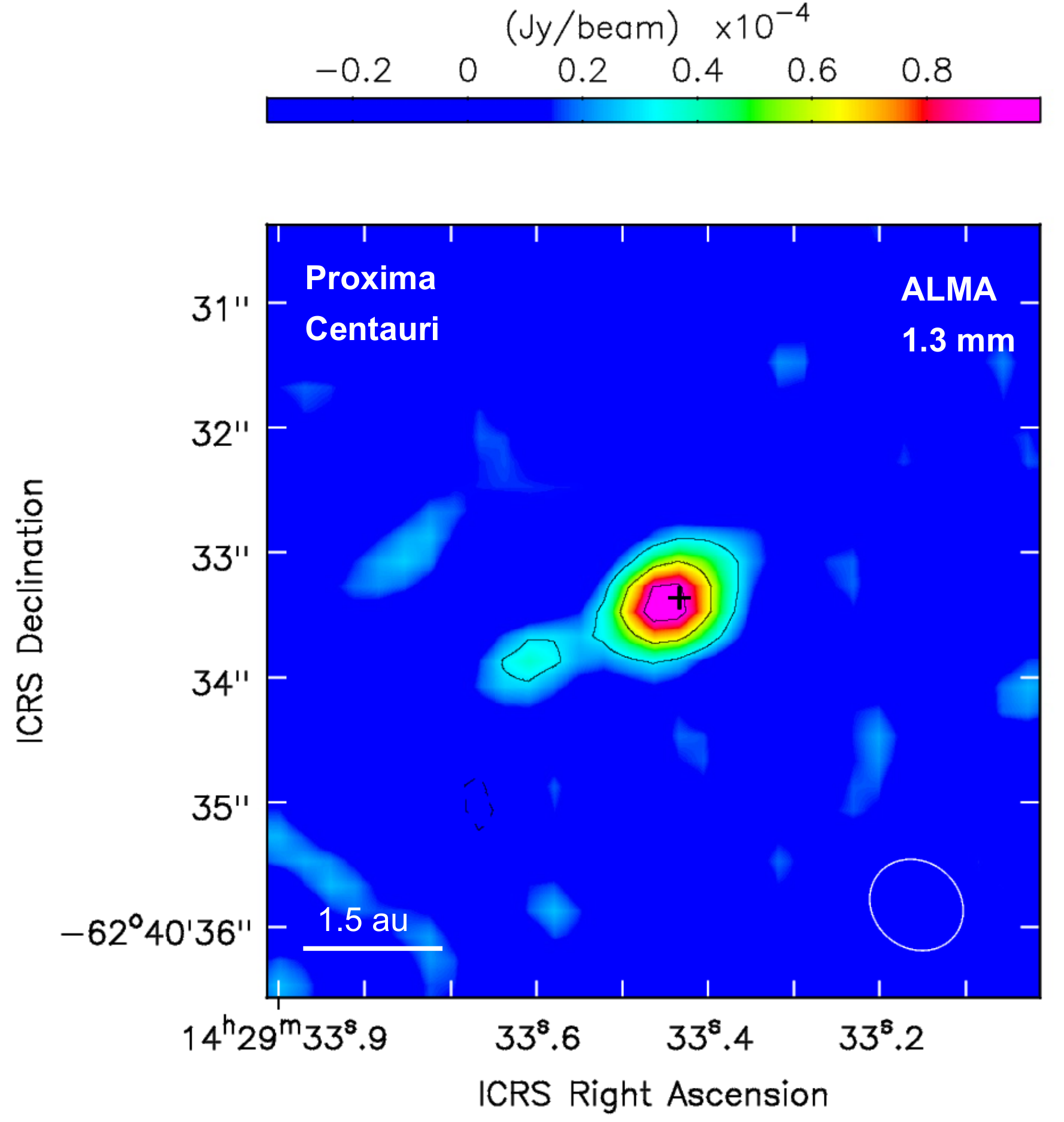}
 \caption{ALMA 12-m array image of the 1.3 mm emission towards 
Proxima Centauri. 
 Contours are $-$3, 3, 6, and 9 times 10 $\mu$Jy beam$^{-1}$, the rms of 
the image.
 The position of the star on 2017 April 25, measured from an optical 
image (error $\simeq$ $0.1''$), is marked with a + sign.
 The synthesized beam ($0.79''\times0.69''$, PA=$50.3\degr$) is shown in 
the bottom-right corner.
 \label{fig:12m}}
\end{center}
 \end{figure}

Figure~\ref{fig:12m} shows the 1.3 mm image obtained with the ALMA 12-m 
array (synthesized beam FWHM$\simeq$$0.7''$). The image shows a main 
central source, whose emission peak is located at ICRS coordinates 
R.A.=$14^h29^m33.445^s\pm0.005^s$, Dec=$-62^\circ40'33.40''\pm0.03''$, 
coinciding within $0.1''$ with the optical position of the star at the 
epoch of observation. The flux density is $106\pm12$ $\mu$Jy and the 
intensity peak is $100\pm12$ $\mu$Jy beam$^{-1}$. The source appears 
marginally elongated (deconvolved size $\sim0.6''$) in a direction with 
PA$\simeq$$130\degr$.

 A secondary source with a flux density of $38\pm10$ $\mu$Jy is 
marginally detected ($4\,\sigma$) at ICRS coordinates 
R.A.=$14^h29^m33.604^s\pm0.011^s$, Dec=$-62^\circ40'33.88''\pm0.07''$, 
at a distance of $1.2''$ and PA=$114\degr$ from the central star.

\section{Results and Discussion} \label{sec:results}

\subsection{The Emission from the Star} \label{sec:star}

The star (with a diameter of $\sim 0.001''$) should appear angularly 
unresolved in the ALMA observations. Thus, the observed intensity peak 
in the ALMA 12-m array image (angular resolution $\simeq 0.7''$) sets an 
upper limit of $100\pm10$ $\mu$Jy (Fig.~\ref{fig:12m}) for the stellar 
flux density.

The expected non-thermal emission of the star at 1.3 mm, extrapolated 
from the non-thermal flux density and spectral index measured by Slee et 
al. (2003), as well as our own ATCA measurements between 1 and 3 GHz 
(J.F. G\'omez et al., in prep.), is negligible ($\ll$1 $\mu$Jy).

The expected thermal emission from the star can be obtained from the 
photospheric emission model that fits the overall SED as described by 
Ribas et al. (2017), giving an extrapolated flux density at 1.3 mm of 
$74\pm4\,\mu$Jy. This result is fully consistent with the upper limit 
($<100\,\mu$Jy) provided by our ALMA 12-m array observations 
(Fig.~\ref{fig:12m}), and indicates that $\sim$70-80\% of the flux 
density detected by ACA ($340\pm60\,\mu$Jy; Fig.~\ref{fig:aca}) does not 
arise from the star; therefore, we interpret it as thermal emission from 
circumstellar dust (see below).

\subsection{Dust Around Proxima Centauri} \label{sec:kb}
 
As shown above, the flux density of the unresolved (diameter $\la$ 
$6''$) source detected by ACA largely exceeds what can be accounted for 
by the star, and we interpret the excess of $\sim270\,\mu$Jy as thermal 
emission from dust orbiting the star at radii $r\la3''$ ($r\la4$ au). 
The actual distribution of this dust
 can be further constrained by the ALMA 12-m array data.

If this emission were originated in a compact region (up to a few times 
the solid angle of the ALMA 12-m array synthesized beam), it would have 
been detected with a high S/N in the ALMA image (Fig.~\ref{fig:12m}). 
However, the image only reveals two sources above the $3\,\sigma$ 
threshold within a region of $\sim6''$ in size (similar to the size of 
the ACA synthesized beam), totaling a flux density of $\sim70\,\mu$Jy 
after subtraction of the estimated emission of the star. This implies 
that the remaining $\sim200\,\mu$Jy should be distributed over a solid 
angle $\ga7$ times that of the ALMA beam for its intensity to remain 
below the $3\,\sigma$ threshold of $30\,\mu$Jy beam$^{-1}$ (if a 
fraction of the emission was resolved out by the interferometer, this 
conclusion still holds). This condition requires that a significant part 
of the emission comes from radii larger than $\sim1''$. Thus, with our 
current data we infer that there are $\sim$200 $\mu$Jy of dust emission 
spread over a belt in a range of radii from $\sim1''$ to $\sim3''$, 
corresponding to $\sim$1.3-4 au. A more precise determination of its 
distribution would require additional, more sensitive observations.

While our data are insufficient for a proper modeling, they are 
nevertheless enough to obtain a rough estimate of the masses involved.
 The mass of dust detected by ALMA can be estimated approximately as 
$(m_{\rm dust}/M_\earth)=0.5\, (S_\nu/{\rm Jy})\, (T_d/{\rm K})^{-1}\, 
(d/{\rm pc})^2\, (\nu/{\rm 230\,GHz})^{-2}$, where a dust opacity of 
$\kappa_\nu=2~\rm cm^2~g^{-1}$ has been adopted (e.g., Beckwith et al. 
2000). The dust temperature can be approximated as $(T_d/{\rm K})=278\, 
(L/L_\sun)^{0.25}\, (r/{\rm au})^{-0.5}$ (Wyatt 2008). Thus, for the 
dust observed at $r$$\simeq$1-4 au, we obtain $T_d\simeq40$ K and a dust 
mass $m_d\simeq4\times10^{-6}~M_\earth$.

The 1.3 mm continuum emission traces dust grains with $\mu$m to cm 
sizes. However, this population of small grains results from the 
collisional cascade involving a primordial population of larger bodies 
and planetesimals containing most of the mass. The equilibrium 
size-distribution resulting from this collisional cascade can be 
described by a power-law of index $-$3.5 (Tanaka et al. 1996). Following 
a formulation similar to that of Wyatt \& Dent (2002), it can easily be 
shown that, if the primordial size distribution connects smoothly with 
the collisional cascade distribution, the total mass can be approximated 
by $m_{\rm tot}\simeq m_{\rm dust} (D_{\rm max}/D_{\rm dust})^{0.5}$, 
where $D_{\rm max}/D_{\rm dust}$ is the ratio between the maximum sizes 
of the population of large bodies and that of the observed dust 
emission. Taking $D_{\rm dust}\simeq 1$ cm, we obtain $m_{\rm 
tot}\simeq2200\, m_{\rm dust}\,(D_{\rm max}/\rm 50\,km)^{0.5}$. 
Therefore, if we integrate up to $D_{\rm max}\simeq50$ km (e.g., Greaves 
et al. 2004; Wyatt et al. 2007a), we obtain a total mass of $m_{\rm 
tot}\simeq10^{-2}~M_\earth$.

This mass is similar to that of the solar Kuiper belt 
($\sim10^{-2}~M_\earth$; Bernstein et al. 2004), which also has a 
similar temperature ($\sim$50 K), but is located at a much larger 
distance (30-50 au) from the Sun. Given the very low luminosity of the 
M-dwarf star Proxima Centauri ($\S1$), one would expect that physical 
conditions similar to those required for the solar Kuiper belt are 
attained at distances much closer to the star. Therefore, we suggest 
that the dust emission in Proxima Centauri, arising at scales $\sim$1-4 
au, is likely tracing a Kuiper belt analog around this star.

Additionally, we note that the central source detected by the ALMA 12-m 
array (Fig.~\ref{fig:12m}) appears marginally elongated along PA 
$\simeq130\degr$ with a flux density of 106 $\mu$Jy and a deconvolved 
size of $\sim$0.8 au. A hint of an excess of emission in the proximity 
of the star and elongation along a similar PA is also observed in images 
made by combining the ACA and ALMA 12-m array data. Thus, considering 
the stellar emission to be 74 $\mu$Jy (see $\S$\ref{sec:star}), it might 
be possible that a small amount of warmer dust, with a flux density of 
$\sim$30 $\mu$Jy is present at distances of $\sim0.4$ au from the star. 
Following the same procedures as above, we estimate a characteristic 
temperature $T_d\simeq90$ K, a dust mass of 
$\sim5.5\times10^{-7}~M_\earth$, and a total mass of 
$\sim10^{-3}~M_\earth$ for this possible hotter component.

\subsection{A Possible Outer Belt at 30 au} \label{sec:outer}

The 1.3 mm ACA image (Fig.~\ref{fig:aca}) does not show direct evidence 
for dust structures other than the compact central source, but it shows 
a number of weak emission peaks that could be part of a larger 
structure. Since a tilted circular belt would appear as an ellipse on 
the sky, we performed averaging of the observed emission over elliptical 
annuli centered on the star to increase the S/N. When azimuthally 
averaging the intensity in annuli, the S/N of the intensity profile at a 
given radius improves by a factor equal to the square root of the ratio 
between the length of the annulus and the beam diameter.
 This approach has been successfully used to infer the presence of rings 
and gaps in protoplanetary disks around young stars (Osorio et al. 2014; 
Macias et al. 2017), as well as in debris disks around old stars (Marino 
et al. 2017).

\begin{figure}[htb]
\begin{center}
\epsscale{0.8}
\plotone{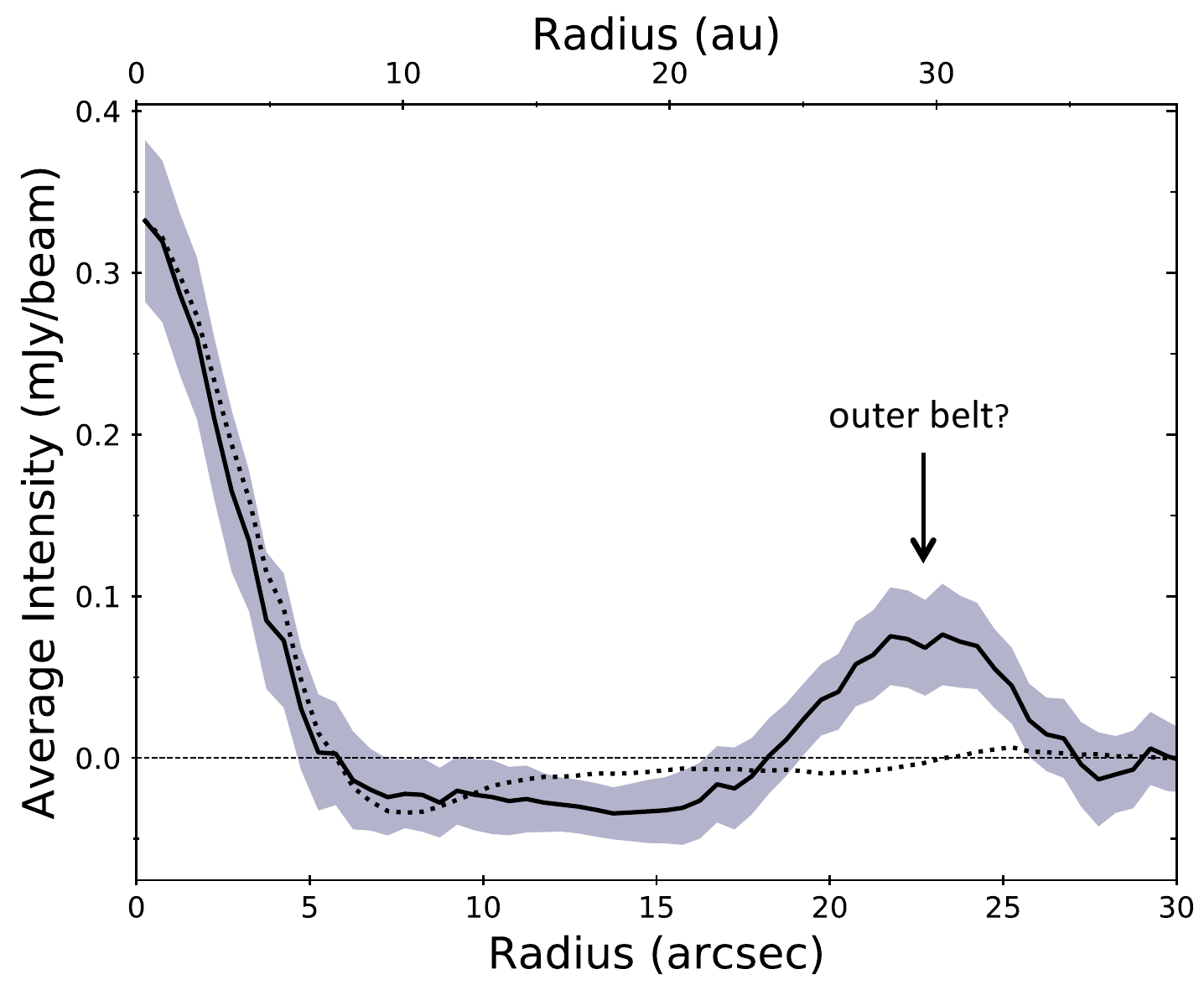}
 \caption{Azimuthally averaged radial intensity profile (solid line) of 
the primary-beam corrected ACA image. Averaging has been performed over 
elliptical annuli, centered on the position of the star and with the 
major axis along PA=$140\degr$ (see Fig.~\ref{fig:aca}), corresponding 
to the projection on the plane of the sky of circular annuli with an 
inclination angle $i=45\degr$. The plot shows the compact central source 
and a possible outer belt at a (deprojected) radius of $\sim23''$ 
($\sim$30 au).
 The gray area illustrates the 1\,$\sigma$ uncertainty, calculated from 
the standard deviation of the observed intensity within each annulus, which 
gives an upper limit for the uncertainty.
 The synthesized beam profile (dotted line) averaged over the same 
region is also plotted as a reference.
 \label{fig:profile}}
\end{center}
 \end{figure}

The radial intensity profiles obtained in this way suggest an emission 
peak around a radius of $\sim23''$ ($r\simeq30$ au). We analyzed a grid 
of different position angles and eccentricities of the ellipses. The 
peak appears as better defined and stronger for a combination of the 
ellipse PA $\simeq140\degr$ (major axis) and an eccentricity 
corresponding to an inclination angle $i\simeq45\degr$ (see 
Figs.~\ref{fig:profile} and \ref{fig:aca}). Deviations from this 
combination of parameters produce a progressive vanishing of the 
feature. This feature is only detected in the ACA image because it 
appears at large distances from the center where the response of the 
ALMA 12-m primary beam is very low because of its smaller FWHM (see 
$\S$\ref{sec:obs}). If this belt proves to be real, it would provide a 
good estimate of the orientation of the orbital plane of the Proxima 
Centauri planetary system. If coplanarity is assumed, the true mass of 
the planet Proxima b (Anglada-Escud\'e et al. 2016) would be 
$\sim1.8\,M_\earth$. Interestingly, the PA of this outer belt is similar 
to that found for the elongated central source ($\sim130\degr$; 
Fig.~\ref{fig:12m}), supporting the reality of both structures, and 
suggesting a similar orientation of the system at both small and large 
scales. However, we should emphasize that the detection of this outer 
belt is marginal, and it should be confirmed with additional 
observations of higher sensitivity. We note that the $45\degr$ 
inclination of this tentative outer belt differs significantly from the 
$108\degr$ inclination of the orbit of Proxima Centauri around alpha 
Centauri AB (Kervella et al. 2017). However, we do not consider these 
results to be in conflict since, in general, orbital motions in a 
hierarchical triple system are not expected to be coplanar (e.g., 
Mu\~noz \& Lai 2015). Indeed, the orbit of the alpha Centauri AB binary 
also has a significantly different inclination ($79\degr$; Kervella et 
al. 2016).

\begin{figure}[htb]
\begin{center}
\epsscale{0.9}
\plotone{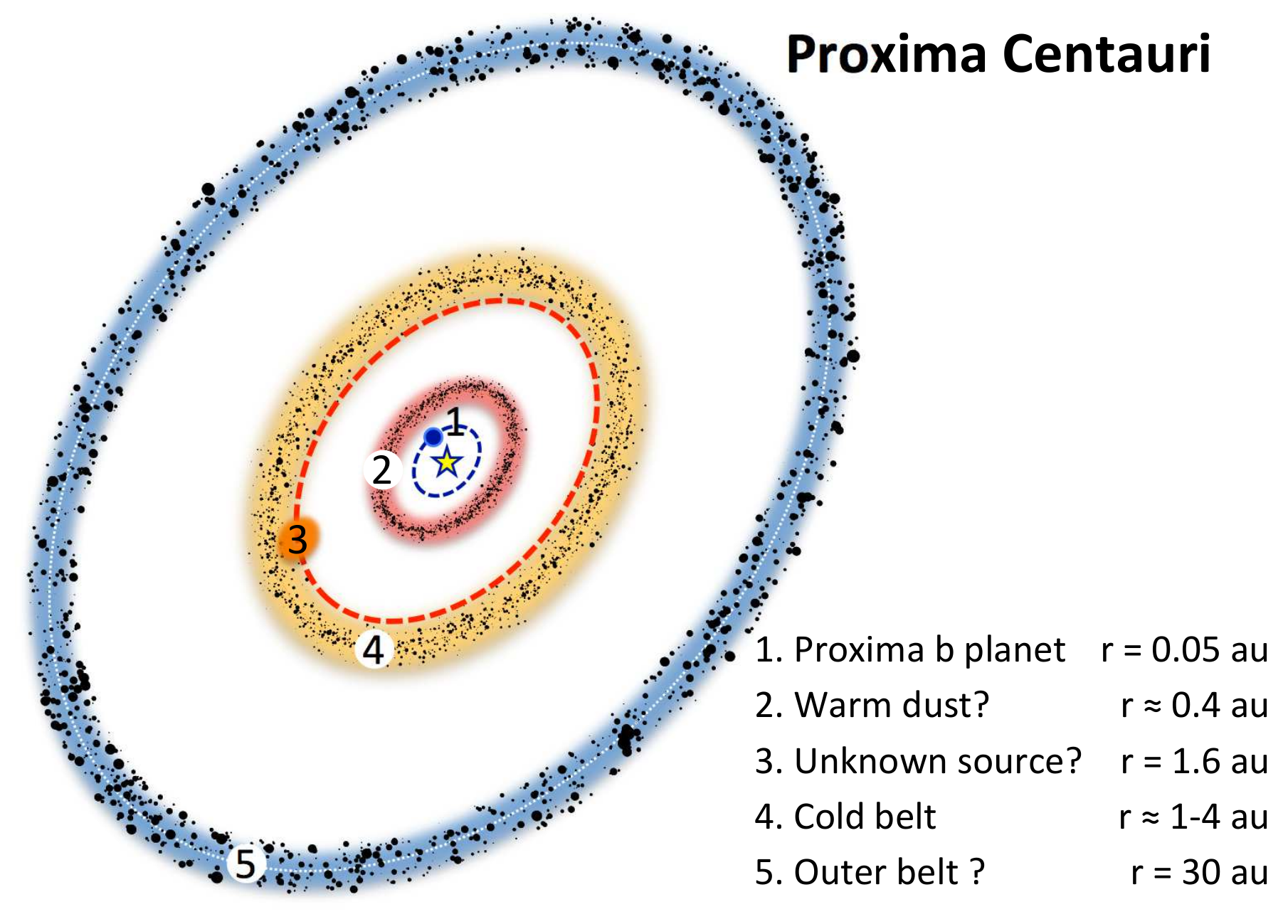}
 \caption{Sketch (not to scale) of the proposed components in the 
Proxima Centauri planetary system. Question marks indicate marginally 
detected features.
 \label{fig:cartoon}}
\end{center}
 \end{figure}

Thus, it is possible that Proxima Centauri is surrounded by several 
belts of dust (see Fig.~\ref{fig:cartoon}), one or several close to the 
star ($r\la4$ au), and another one (to be confirmed) at a large radius 
($r\simeq30$ au). For this outer belt the average intensity is $\sim 
76~\mu$Jy~beam$^{-1}$ (Fig.~\ref{fig:profile}), and we estimate a total 
flux density of $\sim$ 1.7 mJy. The presence of such a distant belt in 
the very low luminosity star Proxima Centauri is challenging. It would 
be extremely cold (with $T_d\simeq10$ K, if the same temperature law is 
assumed) and its flux density would lead to a dust mass of 
$1.4\times10^{-4}~M_\earth$, corresponding to a total mass (including 
large bodies) of $0.33~M_\earth$, if the same assumptions as in 
$\S$\ref{sec:kb} are made, but see Krivov et al. 2013. This mass is much 
larger than that of the solar Kuiper belt ($\sim10^{-2}\, M_\earth$; 
Bernstein et al. 2004), and so far, there is no known analog in the 
solar system. We note, however, that {\em Herschel} revealed a new class 
of very cold debris disks around some mature solar-type stars (Eiroa et 
al. 2011), whose origin (still unexplained) could share some 
similarities with our proposed 30 au belt in Proxima Centauri.

\subsection{A Ring around an Undiscovered Planet at $\sim$1.6 au?} 
\label{sec:planet}

As noted in $\S$\ref{sec:obs}, our images show a marginally detected 
($\sim 40~\mu$Jy at $4\,\sigma$) compact source of 1.3 mm emission at a 
projected distance of $\sim 1.2''$ SE from the star (see 
Fig.~\ref{fig:12m}). This source is very intriguing.
 We can discard beam or cleaning artifacts, given the low beam sidelobes 
and high quality of the image. However, the significance of the source 
is marginal, and with the current data we cannot rule out the 
possibility that this component is just a noise peak.

If the source was real we consider several possibilities. The source 
could be a background galaxy (Smail et al. 1997; Fujimoto et al. 2016). 
The probability of finding a source like this within $1.2''$ of Proxima 
Centauri is small ($\la10^{-2}$, according to the source counts of 
Fujimoto et al. 2016) but it cannot be completely discarded. Given the 
large proper motions of the star, a second-epoch observation would 
easily reveal whether this source moves together with the star or if it 
is a background static source.
 A substellar companion with a temperature of the order of 1000 K could 
also produce the observed emission. However, according to our 
calculations, such an object would produce a detectable signature in the 
radial velocity (RV) of the star that has not been observed.
 A transient event, such as the collision between large bodies (Wyatt et 
al. 2007b) might produce a cloud of dust with properties similar to the 
observed source. However, observation of such an episodic event seems 
unlikely in this old star.
 The source might be tracing a cloud of dust orbiting in the proximity 
of the Lagrange points of an still undetected planet, in a way similar 
to the Trojan minor planets in our solar system. However, Trojan clouds 
are located at the L$_4$ and L$_5$ Lagrange points $\sim60\degr$ ahead 
and behind the larger body.
 If the 140$\degr$ position angle of the $\sim30$ au tilted belt is 
significant, and the inner dust disks and potential planetary orbits 
share the same inclination, their brightest parts would be approximately 
SE and NW, so we might be seeing the more favorably placed Trojan cloud, 
or possibly just the brightest part of an uneven disk which is mostly 
just below our detection threshold at the higher resolution.

Finally, an exciting alternative scenario is that the source traces a 
ring of dust surrounding an as yet undiscovered giant planet orbiting at 
a (projected) distance of 1.6 au (orbital period $\ga$ 5.8 yr). By 
analogy with the rings of Saturn we expect a power-law distribution with 
an index of $-3.5$ and a maximum particle size of 5-10 m (Zebker et al. 
1985; Brilliantov et al. 2015), resulting in a total mass of 
$\sim10^{-5}~M_\earth$ for such a planetary ring. Theoretical arguments 
(Charnoz et al. 2017) suggest that evolved planetary rings have a mass 
$\sim10^{-7}$ times the mass of the planet. Thus, under this scenario, 
we would expect a planet of mass $\sim100~M_\earth$, the mass of Saturn, 
to account for the observed 1.3 mm emission. No clear RV signal that 
would indicate such a planet is present in the data of the long-term 
monitoring of the star. Further observations are being undertaken to 
confirm, or rule out this intriguing possibility. At any rate, our study 
shows that ALMA provides already the necessary sensitivity and 
resolution to detect rings around exoplanets in alpha Centauri, and 
perhaps in other nearby stars.

\acknowledgments

This paper makes use of the following ALMA data: 
ADS/JAO.ALMA\#2016.A.00013.S. ALMA is a partnership of ESO (representing 
its member states), NSF (USA) and NINS (Japan), together with NRC 
(Canada) and NSC and ASIAA (Taiwan) and KASI (Republic of Korea), in 
cooperation with the Republic of Chile. The Joint ALMA Observatory is 
operated by ESO, AUI/NRAO and NAOJ. This work has been partially 
supported by MINECO (Spain) grants co-funded with FEDER. I.J.-S. 
acknowledges financial support from the STFC through an Ernest 
Rutherford Fellowship. J.S.J. acknowledges support from Fondecyt and 
from CATA-Basal.
 Z.M.B. acknowledges partial support from ALMA-CONICYT FUND.
We are deeply indebted to the ALMA Observatory staff for their efficient handling of the observations and specially to the UK ARC Node, and data analysts in North American and European ALMA Regional Centers for their fast reaction in the data reduction process.



\end{document}